\documentstyle[prl,aps,graphics,epsfig]{revtex}
\renewcommand{\theequation}{{\rm\arabic{section}.\arabic{equation}}}

\begin{document}
\begin{center}
\bigskip\bigskip\bigskip
{\Huge On the Measurement of Qubits}\\
\bigskip\bigskip\bigskip
{\LARGE Daniel F. V. James$^{(1)}$
\footnote{Corresponding author: Mail stop B-283,
Los Alamos National Laboratory, Los Alamos NM  87545, USA;
tel: (505) 667-5436; FAX: (505) 665-1931; e-mail: dfvj@lanl.gov.},
Paul G. Kwiat$^{(2,3)}$,\\
\bigskip
William J. Munro$^{(4,5)}$ and Andrew G. White$^{(2,4)}$}\\
\bigskip\bigskip
{\large (1) Theoretical Division T-4, Los Alamos National Laboratory,\\
Los Alamos, New Mexico 87545, USA.\\
\bigskip
(2) Physics Division P-23, Los Alamos National Laboratory,\\
Los Alamos, New Mexico 87545, USA.\\
\bigskip
(3) Dept. of Physics, University of Illinois,\\
Urbana-Champaign, Illinois  61801, USA.\\
\bigskip
(4) Department of Physics, University of Queensland,\\
Brisbane, Queensland 4072, AUSTRALIA\\
\bigskip
(5) Hewlett-Packard Laboratories, Filton Road, \\
Stoke Gifford, Bristol BS34 8QZ, UNITED KINGDOM}\\
\bigskip\bigskip
To be submitted to {\em Phys Rev A} \\
\bigskip
\bigskip
Date: \today

\bigskip\bigskip
{\large {\bf Abstract}}\\
\end{center}
\begin{quote}
We describe in detail the theory underpinning the measurement of
density matrices of a pair of quantum two-level
systems (``qubits'').  Our particular emphasis is on qubits realized
by the two polarization degrees of
freedom of a pair of entangled photons generated in a down-conversion
experiment; however the discussion applies in general,
regardless of the actual physical realization.
Two techniques are discussed, namely a tomographic
reconstruction (in which the density matrix is linearly related to
a set of measured quantities) and a maximum likelihood technique
which requires numerical optimization (but has the advantage
of producing density matrices which are always non-negative definite).
In addition a detailed error analysis is presented, allowing
errors in quantities derived from the density matrix, such as the
entropy or entanglement of formation, to be estimated.  Examples
based on down-conversion experiments are used to illustrate our
results.
\end{quote}
\begin{center}
\bigskip
LA-UR-01-1143
\end{center}

\newpage
\section{Introduction}
\setcounter{equation}{0}
The ability to create, manipulate and characterize
quantum states is becoming an increasingly important area of
physical research, with implications for areas of
technology such as quantum computing, quantum cryptography and
communications.
With a series of measurements on a large enough number of 
identically prepared copies of an
quantum system, one can infer,to a reasonable approximation,
the quantum state of the system.
Arguably, the first such experimental technique for determining the
state of quantum system was devised by
George Stokes in 1852 \cite{Stokes}.  His famous four parameters
allow an experimenter to determine uniquely the polarization state
of a light beam.  With the insight provided by nearly 150 years
of progress in optical physics, we can consider coherent
light beams to be an ensemble of two-level quantum mechanical
systems, the two levels being the two polarization degrees of freedom
of the photons; the Stokes parameters allow one to determine the
density matrix describing this ensemble.
More recently, experimental techniques for the measurement of the
more subtle  quantum properties of light have been the subject of intensive
investigation (see ref.\cite{leonhardt} for a comprehensive and erudite
exposition of this subject).  In various experimental circumstances is 
has been found reasonably straightforward
to devise a simple linear tomographic technique in which the density matrix
(or Wigner function) of a quantum state is found from a linear transformation
of experimental data. However, there is one important drawback to this
method, in that the recovered state might not correspond to a physical
state because of experimental noise.  For example, density matrices for
any quantum state must be Hermitian, positive semi-definite matrices
with unit trace.  The tomographically measured matrices often fail to
be positive semi-definite, especially when measuring low-entropy 
states. To avoid this problem the ``maximum
likelihood'' tomographic approach to the
estimation of quantum states has been developed
\cite{hradil1,tan,banaszek,hradil,rehacek}.  In this approach
the density matrix that is ``mostly likely'' to have produced a
measured data set is determined by numerical optimization.

In the past decade several groups have successfully employed
tomographic techniques for
the measurement of quantum mechanical systems. 
In 1990 Risley et al. at North Carolina State University reported the measurement
of the density matrix for the nine sublevels of the $n=3$ level of 
hydrogen atoms formed following collision between ${\rm H}^{+}$ ions and 
He atoms, in conditions of high symmetry which simplified the
tomographic problem \cite{ashburn}.  
Since then, in 1993 Smithey et al. (University of Oregon),
made a homodyne measurement of the Wigner function of a single mode of light \cite{Raymer}.  
Other explorations of the quantum states of single mode light fields
have been made by Mlynek et al. (University of Konstanz, Germany) \cite{Mlynek}  
and Bachor et al. (Australian National University) \cite{bachor}.
Other quantum systems whose density matrices have been investigated 
experimentally include the vibrations of molecules \cite{Walmsley}, 
the motion ions and atoms \cite{Wineland,Kurtseifer} and 
the internal angular momentum quantum state of the $F=4$ ground state of a cesium
atom \cite{Jessen}.  The quantum states of multiple spin-half
nuclei have been measured in the high-temperature regime using NMR
techniques \cite{Chuang}, albeit in systems of such high entropy that
the creation of entangled states is necessarily precluded
\cite{Braunstein}.  The measurement of the quantum state of entangled
qubit pairs, realized using the polarization degrees of freedom of a pair of
photons created in a parametric down-conversion experiment was
reported by us recently \cite{white}.

In this paper we will examine techniques in detail for quantum state
measurement as it applies to multiple correlated two-level quantum
mechanical systems (or ``qubits'' in the terminology of quantum
information).  Our particular emphasis is qubits realized via the two
polarization degrees of freedom of photons, data from which we use
to illustrate our results.  However, these techniques are readily
applicable to other technologies proposed for creating entangled
states of pairs of two-level systems. Because of the central
importance of qubit systems to the emergent discipline of quantum
computation, a thorough explanation of the techniques needed to
characterize the qubit states will be of relevance to workers in the
various diverse experimental fields currently under consideration
for quantum computation technology \cite{fordphys}.  This paper is
organized as follows: In Section
II we explore the analogy with the Stokes parameters, and how they
lead naturally to a scheme for measurement of an arbitrary number
of two-level systems.  In Section III, we discuss the measurement
of a pair of qubits in more detail, presenting the validity condition
for an arbitrary measurement scheme and introducing the set of 16
measurements employed in our experiments.  Section IV deals with our
method for maximum likelihood reconstruction and in Section V we
demonstrate how to calculate the
errors in such measurements, and how these errors propagate to
quantities calculated from the density matrix.


\section{The Stokes parameters and quantum state tomography}
As mentioned above, there is a direct analogy between the
measurement of the polarization state of a light beam and
the measurement of the density matrix of an ensemble of two-level
quantum mechanical systems.  Here we explore this analogy in more
detail.

\subsection{Single qubit tomography}
The Stokes parameters are defined from a set of four
intensity measurements\cite{HZ} :
(i) with a filter that transmits 50$\%$ of the incident
radiation, regardless of its polarization; (ii) with a polarizer
that transmits only horizontally polarized light; (iii) with a polarizer
that transmits only light polarized at $45^{o}$ to the horizontal; and
(iv) with a polarizer
that transmits only right circularly polarized light.  The number
of photons counted by a detector, which is proportional
to the classical intensity, in these four experiments are as follows:
\begin{eqnarray}
n_{0} &=& \frac{{\cal N}}{2} \left( \langle H |\hat{\rho} | H \rangle
+\langle V |\hat{\rho} | V \rangle\right)
=\frac{{\cal N}}{2} \left( \langle R |\hat{\rho} | R \rangle
+\langle L |\hat{\rho} | L \rangle\right)
\nonumber \\
n_{1} &=& {\cal N} \left( \langle H |\hat{\rho} | H \rangle \right)
\nonumber \\
&=& \frac{{\cal N}}{2} \left(
  \langle R |\hat{\rho} | R \rangle
+\langle R |\hat{\rho} | L\rangle
+\langle L |\hat{\rho} | R \rangle
+\langle L |\hat{\rho} | L \rangle\right)
\nonumber \\
n_{2} &=& {\cal N} \left( \langle \bar{D} |\hat{\rho} | \bar{D} \rangle \right)
\nonumber \\
&=& \frac{{\cal N}}{2} \left(
  \langle R |\hat{\rho} | R \rangle
+\langle L |\hat{\rho} | L \rangle
-i\langle L |\hat{\rho} | R \rangle
+i\langle R |\hat{\rho} | L \rangle\right)
\nonumber \\
n_{3} &=& {\cal N} \left( \langle R |\hat{\rho} | R \rangle \right)
\nonumber \\
&&
\end{eqnarray}
Here $| H \rangle $,  $| V \rangle $,  $| \bar{D} \rangle = \left(| H 
\rangle -| V \rangle \right)/\sqrt{2} = \exp(i\pi/4)\left(| R 
\rangle +i| L \rangle \right)/\sqrt{2}$ and  
$| R \rangle = \left(| H \rangle - i| V
\rangle \right)/\sqrt{2} $
are the kets representing qubits polarized in the  linear horizontal,
linear vertical, linear diagonal ($45^{o}$) and right-circular senses
respectively, $\hat{\rho}$ is the $(2 \times 2)$ density matrix for
the polarization degrees of the light (or, for a two-level quantum
system) and
${\cal N}$ is  constant dependent on the detector efficiency and light
intensity.  The {\em Stokes parameters}, which fully characterize the
polarization state of the light, are then defined by
\begin{eqnarray}
{\cal S}_{0} &\equiv& 2 n_{0}
={\cal N} \left(
\langle R |\hat{\rho} | R \rangle +\langle L |\hat{\rho} | L \rangle
\right)
\nonumber \\
{\cal S}_{1} &\equiv& 2 \left(n_{1}-n_{0}\right)
={\cal N} \left(
\langle R |\hat{\rho} | L \rangle +\langle L |\hat{\rho} | R \rangle
\right)
\nonumber \\
{\cal S}_{2} &\equiv& 2 \left(n_{2}-n_{0}\right)
={\cal N} i\left(
\langle R |\hat{\rho} | L \rangle -\langle L |\hat{\rho} | R \rangle
\right)
\nonumber \\
{\cal S}_{3} &\equiv& 2 \left(n_{3}-n_{0}\right)
= {\cal N} \left(
\langle R |\hat{\rho} | R \rangle -\langle L |\hat{\rho} | L \rangle
\right)
\nonumber . \\
&&
\label{repulse}
\end{eqnarray}

We can now relate the Stokes parameters to the density matrix $\hat{\rho}$ by 
the formula
\begin{equation}
    \hat{\rho} = \frac{1}{2}\sum_{i=0}^{3} \frac{{\cal S}_{i}}{{\cal S}_0}
    \hat{\sigma}_{i},
\end{equation}
where  $\hat{\sigma}_{0} = |R\rangle\langle R|+|L\rangle\langle L|$
is the single qubit identity operator and
$\hat{\sigma}_{1} = |R \rangle\langle L|+|L \rangle\langle R|$,
$\hat{\sigma}_{2} = i(|L \rangle\langle R|-|R \rangle\langle L|$
and $\hat{\sigma}_{3} = |R \rangle\langle R|-|L \rangle\langle L|$
are the Pauli spin operators. Thus the measurement of the
Stokes parameters can be considered equivalent to a tomographic
measurement of the density matrix of an ensemble of single
qubits.

\subsection{Multiple beam Stokes parameters: multiple qubit tomography}

The generalization of the Stokes scheme to measure the state of
multiple photon beams (or multiple qubits) is reasonably straightforward.
One should, however, be
aware that importance differences exist between the one-photon and
the multiple photon cases.  Single photons, at least in the current
context, can be described in a
purely {\em classical} manner, and the density matrix can be related to
the purely classical concept of the coherency matrix \cite{wolf}.
For multiple photon one has the possibility of
non-classical correlations occurring, with quintessentially
quantum-mechanical phenomena such as entanglement being present.  We
will return to the concept of entanglement and how it may be measured
later in this paper.

An n-qubit state is characterized by a density matrix which may be
written as follows:
\begin{equation}
    \hat{\rho} = \frac{1}{2^{n}} \sum_{i_{1}, i_{2},\ldots i_{n} = 0}^{3}
    r_{i_{1}, i_{2},\ldots i_{n}} \hat{\sigma}_{i_{1}}\otimes
    \hat{\sigma}_{i_{2}}\otimes\ldots\otimes \hat{\sigma}_{i_{n}},
    \label{warspite}
\end{equation}
where the $4^{n}$ parameters $r_{i_{1}, i_{2},\ldots i_{n}}$ are real
numbers.  The normalization property of the density matrices requires
that $r_{0, 0,\ldots 0} = 1$, and so the density matrix is specified
by $4^{n}-1$ real parameters.  The symbol $\otimes$ represents the
tensor product between operators acting on the Hilbert spaces
associated with the separate qubits.

As Stokes showed, the state of a single qubit can be determined by
taking a set of four projection measurements which are represented by
the four operators $\hat{\mu}_{0} = |H\rangle\langle H|+|V\rangle\langle
V|$, $\hat{\mu}_{1} = |H\rangle\langle H|$, $\hat{\mu}_{2} = |\bar{D}\rangle\langle
\bar{D}|$, $\hat{\mu}_{3} = |R\rangle\langle R|$.  Similarly, the state of two
qubits can be determined by the set of 16 measurements represented
by the operators $\hat{\mu}_{i} \otimes \hat{\mu}_{j}$ ($i,j =
0,1,2,3$).  More generally the state of an n-qubit system can be
determined by $4^{n}$ measurements given by the operators $\hat{\mu}_{i_{1}} \otimes
\hat{\mu}_{i_{2}} \otimes \ldots \otimes \hat{\mu}_{i_{n}} $ ($i_{k} =
0,1,2,3$) and ($ k = 1, 2, \ldots n$).  This `tree' structure for
multi-qubit measurement is illustrated in fig.1.

The proof of this conjecture is reasonably straightforward.  The
outcome of a measurement is given by the formula
\begin{equation}
n = {\cal N}{\rm Tr} \left\{ \hat{\rho}\hat{\mu} \right\},
\end{equation}
where $\hat{\rho}$ is the density matrix, $\hat{\mu}$ is the
measurement operator and $\cal{N}$ is a constant of proportionality
which can be determined from the data. Thus in our n-qubit case, the outcomes
of the various measurement are
\begin{equation}
n_{i_{1}, i_{2},\ldots i_{n}} = {\cal N}{\rm Tr} \left\{
\hat{\rho}\left(\hat{\mu}_{i_{1}}\otimes
    \hat{\mu}_{i_{2}}\otimes\ldots\otimes
    \hat{\mu}_{i_{n}}\right) \right\}.
\end{equation}
Substituting from eq.(\ref{warspite}) we obtain
\begin{equation}
n_{i_{1}, i_{2},\ldots i_{n}} = \frac{{\cal N}}{2^{n}} \sum_{j_{1},
j_{2},\ldots j_{n} = 0}^{3}
{\rm Tr} \left\{\hat{\mu}_{i_{1}} \hat{\sigma}_{j_{1}} \right\}
{\rm Tr} \left\{\hat{\mu}_{i_{2}} \hat{\sigma}_{j_{2}} \right\}
\ldots
{\rm Tr} \left\{\hat{\mu}_{i_{n}} \hat{\sigma}_{j_{n}} \right\}
r_{i_{1}, i_{2},\ldots i_{n}}.
\label{valiant}
\end{equation}
As can be easily verified, the single qubit measurement operators $\hat{\mu}_{i}$ are linear
combinations of the Pauli operators $\hat{\sigma}_{j}$, i.e.
$\hat{\mu}_{i} = \sum_{j=0}^{3} \Upsilon_{ij}\hat{\sigma}_{j}$,
where $\Upsilon_{ij}$ are the elements of the matrix
\begin{equation}
    \Upsilon = \left( \begin{array}{cccc}
      1  &   0  &   0  &  0    \\
     1/2 &  1/2 &   0  &  0    \\
     1/2 &   0  &  1/2 &  0    \\
     1/2 &   0  &   0  &  1/2  \end{array}
     \right).
\end{equation}
Further, we have the relation ${\rm Tr} \left\{\hat{\sigma}_{i}
\hat{\sigma}_{j} \right\} = 2 \delta_{ij}$ (where $\delta_{ij}$
is the Kronecker delta).  Hence eq.(\ref{valiant}) becomes
\begin{equation}
n_{i_{1}, i_{2},\ldots i_{n}} = {\cal N} \sum_{j_{1},
j_{2},\ldots j_{n} = 0}^{3}
\Upsilon_{i_{1}j_{1}}
\Upsilon_{i_{2}j_{2}}
\ldots
\Upsilon_{i_{n}j_{n}}
r_{i_{1}, i_{2},\ldots i_{n}}.
\label{barham}
\end{equation}
Introducing the left-inverse of the matrix $\Upsilon$, defined so
that $\sum_{k=0}^{3} (\Upsilon^{-1})_{ik}\Upsilon_{kj} = \delta_{ij}$
and whose elements are
\begin{equation}
    \Upsilon^{-1} = \left( \begin{array}{cccc}
      1  &   0  &   0  &  0   \\
     -1  &   2  &   0  &  0 \\
     -1  &   0  &   2  &  0   \\
     -1 &    0  &   0  &  2 \end{array}
     \right),
\end{equation}
we can find a formula for the parameters $r_{i_{1}, i_{2},\ldots
i_{n}}$ in terms of the measured quantities $n_{i_{1}, i_{2},\ldots
i_{n}}$,
viz.:
\begin{eqnarray}
    {\cal N} r_{i_{1}, i_{2},\ldots i_{n}} &=&  \sum_{j_{1}, j_{2},\ldots j_{n}
= 0}^{3}
    \left(\Upsilon^{-1}\right)_{i_{1}j_{1}}
    \left(\Upsilon^{-1}\right)_{i_{2}j_{2}}
    \ldots
    \left(\Upsilon^{-1}\right)_{i_{n}j_{n}}
    n_{i_{1}, i_{2},\ldots i_{n}} \nonumber \\
    &\equiv& {\cal S}_{i_{1}, i_{2},\ldots i_{n}}. \label{malaya}
\end{eqnarray}
In eq.(\ref{malaya}) we have introduced the n-photon Stokes
parameter ${\cal S}_{i_{1}, i_{2},\ldots i_{n}}$, defined in
an analogous manner to the single photon Stokes parameters
give in eq.(\ref{repulse}).

Since, as already noted, $r_{0,0,\ldots 0} = 1$, we can make the
identification ${\cal S}_{0,0,\ldots0} = {\cal N}$, and so the
density matrix for the n-qubit system can be written in terms of
the Stokes parameters as follows:
\begin{equation}
    \hat{\rho} = \frac{1}{2^{n}} \sum_{i_{1}, i_{2},\ldots i_{n} = 0}^{3}
    \frac{ {\cal S}_{i_{1}, i_{2},\ldots i_{n}}}{{\cal S}_{0,0,\ldots0}}
\hat{\sigma}_{i_{1}}\otimes
    \hat{\sigma}_{i_{2}}\otimes\ldots\otimes \hat{\sigma}_{i_{n}}.
    \label{queenelizabeth}
\end{equation}
This is a recipe for measurement of the density matrices which,
assuming perfect experimental conditions and the complete absence of
noise, will always work.  
It is important to realize that the set of four
Stokes measurements $\left\{ \hat{\mu}_{0}, \hat{\mu}_{1}, 
\hat{\mu}_{2}, \hat{\mu}_{3} \right\}$ are not unique:
there may be circumstances in which it
is more convenient to use some other set, which are equivalent.
A more typical set, at least in optical  experiments, is 
$\hat{\mu}^{\prime}_{0} = |H\rangle\langle H|$, 
$\hat{\mu}^{\prime}_{1} = |V\rangle\langle V|$, $\hat{\mu}^{\prime}_{2} = |D\rangle\langle
D|$, $\hat{\mu}^{\prime}_{3} = |R\rangle\langle R|$.

In the following section we will explore more general schemes for the
measurement of two qubits, starting with a discussion, in some detail,
of how the measurements are actually performed.

\section{Generalized Tomographic Reconstruction of the Polarization State of Two
Photons}
\setcounter{equation}{0}

\subsection{Experimental set-up}
The experimental arrangement used in our experiments is shown
schematically in Fig.1.  An optical system consisting of
lasers, polarization elements and non-linear optical crystals
(and collectively characterized for the purposes of this paper as a
``black-box'',) is
used to generate pairs of qubits in an almost arbitrary quantum
state of their polarization degrees of freedom.
A full description of this optical system and how such quantum states
can be prepared can be found in ref.\cite{blackbox1,blackbox2,berglund}
\footnote{It is important to realize that the entangled
photon pairs are 
produced in a {\em non-deterministic} manner: one cannot 
specify with certainly when a photon pair will be emitted; indeed there
is a small probability of generating four, or six or higher numbers of 
photons.  Thus we can only {\em post-selectively} generate entangled 
photon pairs: i.e. one only knows that the state was created after if
has been measured.}
The output of the black box consists of a pair of beams of light,
whose quanta can be measured by means of photo-detectors.  To
project the light beams onto a polarization state of the experimenter's
choosing, three optical elements are placed in the beam in front of
each detector: a polarizer (which transmits only vertically polarized
light), a quarter-wave plate and a half-wave plate.  The angles of
the fast axes of both of the waveplates can be set arbitrarily,
allowing the $|V\rangle$ projection
state fixed by the polarizer to be rotated into any polarization state
that the experimenter may wish.

Using the Jones calculus notation, with the following convention,
\begin{equation}
\left( \begin{array}{c} 0\\ 1\\ \end{array}\right) = |\rm{V}\rangle,
\,\,\,
\left( \begin{array}{c} 1\\ 0\\ \end{array}\right) = |\rm{H}\rangle,
\end{equation}
where $|\rm{V}\rangle$ ($|\rm{H}\rangle$) is the ket for a vertically
(horizontally) polarized beam, the effect of quarter and half wave
plates whose fast axes are at angles $q$ and $h$ with respect to the 
vertical axis, respectively, are given by the 2 $\times$ 2 matrices
\begin{eqnarray}
\hat{U}_{QWP}(q)&=&\frac{1}{\sqrt{2}}\left( \begin{array}{cc} 
i-\cos(2q) &  \sin(2q)\\
\sin(2q)  & i+\cos(2q)\\ 
\end{array}\right) ,\nonumber \\
\hat{U}_{HWP}(q)&=& \left( \begin{array}{cc} 
\cos(2h)  &  -\sin (2h)\\
-\sin(2h) & -\cos(2h)\\ 
\end{array}\right) .
\end{eqnarray}
Thus the projection state for the measurement in one of the beams is
given by
\begin{eqnarray}
|\psi^{(1)}_{proj}(h,q)\rangle &=&
\hat{U}_{QWP}(q)\cdot\hat{U}_{HWP}(h)\cdot \left( \begin{array}{c} 0\\
1\\ \end{array}\right) \nonumber \\
&=& a(h,q) |{\rm H}\rangle + b(h,q) |{\rm V}\rangle ,
\end{eqnarray}
where, neglecting an overall phase, the functions $a(h,q)$ and $b(h,q)$ are given by
\begin{eqnarray}
a(h,q) &=&
\frac{1}{\sqrt{2}}\left(\sin(2h)-i\sin[2(h-q)]\right), 
\nonumber\\
b(h,q) &=& -\frac{1}{\sqrt{2}}\left(\cos(2h)+i\cos[2(h-q)]\right).
\label{yamato}
\end{eqnarray}

The projection state for the two beams is given by
\begin{eqnarray}
|\psi^{(2)}_{proj}(h_{1},q_{1},h_{2},q_{2})\rangle &=&
|\psi^{(1)}_{proj}(h_{1},q_{1})\rangle
\otimes|\psi^{(1)}_{proj}(h_{2},q_{2})\rangle \nonumber \\
&=&a(h_{1},q_{1}) a(h_{2},q_{2}) |{\rm HH}\rangle +
a(h_{1},q_{1}) b(h_{2},q_{2}) |{\rm HV}\rangle + \nonumber \\
&&b(h_{1},q_{1}) a(h_{2},q_{2}) |{\rm VH}\rangle +
b(h_{1},q_{1}) b(h_{2},q_{2}) |{\rm VV}\rangle.
\label{state}
\end{eqnarray}

We shall denote the projection state corresponding to one particular
set of waveplate angles $\{h_{1,\nu},q_{1,\nu},h_{2,\nu},q_{2,\nu}\}$
\footnote{Here the {\em first} subscript on the waveplate angle refers
one of the two photon beams; the second subscript distinguishes which 
of the sixteen different experimental states is under consideration.}
by the ket $|\psi_{\nu}\rangle$; thus the projection measurement is
represented by the operator $\hat{\mu}_{\nu} =
|\psi_{\nu}\rangle\langle\psi_{\nu}|$.  Consequently, 
the average number of coincidence
counts that will be observed in a given experimental run is
\begin{equation}
n_{\nu} = {\cal N} \langle\psi_{\nu}|\hat{\rho}|\psi_{\nu}\rangle
\label{hornby}
\end{equation}
where  $\hat{\rho}$ is the density matrix describing the ensemble of
qubits, and ${\cal N}$ is a constant dependent on the photon flux and
detector efficiencies.  In what follows, it will be convenient to consider the
quantities $s_{\nu}$ defined by
\begin{equation}
s_{\nu} =  \langle\psi_{\nu}|\hat{\rho}|\psi_{\nu}\rangle.
\label{weetabix}
\end{equation}

\subsection{Tomographically Complete set of Measurements}
In Section II we have given one possible set of projection
measurements $\{|\psi_{\nu}\rangle\ \langle\psi_{\nu}|\}$ which
which uniquely determine the density
matrix $\hat{\rho}$.  However, one can
conceive of situations in which these will not be the most convenient
set of measurements to make.  Here we address the problem of finding
other sets of suitable measurements.  The smallest number of states 
required for such measurements can be
found by a simple argument: there are 15 real unknown
parameters which determine a $4\times 4$ density matrix, plus there is the
single unknown real parameter ${\cal N}$, making a total of 16.  

In order to proceed it is helpful to convert the $4\times 4$ matrix
$\hat{\rho}$ into a 16-dimensional column vector.  To do this we
use a set of 16 linearly independent $4 \times 4$ matrices
$\{\hat{\Gamma}_{\nu}\}$
which have the following mathematical properties:
\begin{eqnarray}
{\rm Tr}\left\{\hat{\Gamma}_{\nu}\cdot\hat{\Gamma}_{\mu}\right\}
&=&\delta_{\nu,\mu} \nonumber\\
\hat{A}&=&\sum_{\nu=1}^{16} \hat{\Gamma}_{\nu}
{\rm Tr}\left\{\hat{\Gamma}_{\nu}\cdot\hat{A}\right\} \,\,\, \forall
\hat{A},
\label{gammaconds}
\end{eqnarray}
where $\hat{A}$ is an arbitrary  $4 \times 4$ matrix.
Finding a set of $\hat{\Gamma}_{\nu}$ matrices is in fact reasonably
straightforward: for example, the set of (appropriately normalized)
generators of the Lie algebra $SU(2)\otimes SU(2)$ fulfill the required
criteria (for reference, we list this set in  Appendix A).
These matrices are of course simply a re-labeling of the
two-qubit Pauli matrices $\hat{\sigma}_{i}\otimes \hat{\sigma}_{j}$
($i,j = 0,1,2,3$) discussed above. Using
these matrices the density matrix can be written as
\begin{equation}
\hat{\rho}=\sum_{\nu=1}^{16}\hat{\Gamma}_{\nu} r_{\nu} ,
\label{druitt}
\end{equation}
where $r_{\nu}$ is the $\nu$-th element of a sixteen element
column vector, given by the formula
\begin{equation}
r_{\nu}={\rm Tr}\left\{\hat{\Gamma}_{\nu}\cdot\hat{\rho}\right\}
\end{equation}

Substituting from eq.(\ref{druitt}) into eq.(\ref{hornby}), we obtain
the following linear relationship between the measured coincidence
counts $n_{\nu}$ and the elements of the vector  $r_{\mu}$:
\begin{equation}
n_{\nu}={\cal N} \sum_{\mu=1}^{16}B_{\nu,\mu}r_{\mu}
\label{kosminski}
\end{equation}
where the $16 \times 16$ matrix $B_{\nu,\mu}$ is given by
\begin{equation}
B_{\nu,\mu} = \langle\psi_{\nu}|\hat{\Gamma}_{\mu}|\psi_{\nu}\rangle.
\label{anderson}
\end{equation}
Immediately we find a necessary and sufficient condition for
the completeness of the set of tomographic states
$\{|\psi_{\nu}\rangle\}$: if the matrix $B_{\nu,\mu}$ is non-
singular, then eq.(\ref{kosminski}) can be inverted to give
\begin{equation}
r_{\nu}=({\cal N})^{-1} \sum_{\mu=1}^{16}
\left(B^{-1}\right)_{\nu,\mu}n_{\mu}.
\label{ostrog}
\end{equation}

The set of sixteen tomographic states which we employed
are given in Table 1. They can be shown to satisfy
the condition that $B_{\nu,\mu}$ is non-
singular. By no means are these states unique in this
regard: these were the states chosen principally for
experimental convenience.

These states can be realized by setting specific values of the half-
and quarter-wave plate angles.  The appropriate values of these
angles (measured from the vertical) are given in Table 1.  Note
that overall phase factors do not affect the results of
projection measurements.

Substituting eq.(\ref{ostrog}) into eq.(\ref{druitt}), we find
that
\begin{eqnarray}
\hat{\rho} &=& ({\cal N})^{-1}\sum_{\nu=1}^{16}\hat{M}_{\nu} n_{\nu}
\nonumber \\
&=& \sum_{\nu=1}^{16}\hat{M}_{\nu} s_{\nu} ,
\label{tumblety}
\end{eqnarray}
where the sixteen $4\times 4$ matrices $\hat{M}_{\nu}$ are defined
by
\begin{equation}
\hat{M}_{\nu} = \sum_{\nu=1}^{16}\left(B^{-1}\right)_{\nu,\mu}
\hat{\Gamma}_{\mu}.
\label{chapman}
\end{equation}
The introduction of the $\hat{M}_{\nu}$ matrices allows
a compact form of linear tomographic reconstruction,
eq.(\ref{tumblety}), will be most useful in the error analysis that follows.
These $\hat{M}_{\nu}$ matrices, valid for our set of tomographic states,
are listed in Appendix B, together with some of their
important properties. We can use one of these properties,
eq.(\ref{propone}), to obtain the value of the unknown
quantity ${\cal N}$.  That relationship implies
\begin{equation}
\sum_{\nu}{\rm Tr}\left\{\hat{M}_{\nu}\right\}
|\psi_{\nu} \rangle\langle \psi_{\nu}|\hat{\rho} =
\hat{\rho}.
\end{equation}
Taking the trace of this formula, and multiplying by
${\cal N}$ we obtain:
\begin{equation}
\sum_{\nu}{\rm Tr}\left\{\hat{M}_{\nu}\right\} n_{\nu} = {\cal N}.
\end{equation}
For our set of tomographic states, it can be shown that
\begin{equation}
\sum_{\nu}{\rm Tr}\left\{\hat{M}_{\nu}\right\}=\left\{
\begin{array}{ll}
1 \,\,\,\,& \mbox{if  } \nu = 1,2,3,4 \\
0 & \mbox{if  } \nu = 5,\ldots16 , \\
\end{array}
\right.
\end{equation}
hence the value of the unknown parameter ${\cal N}$  in
our experiments is given by:
\begin{eqnarray}
{\cal N} &=& \sum_{\nu=1}^{4} n_{\nu} \nonumber \\
&=& {\cal N} \left( \langle HH|\hat{\rho}|HH\rangle +
\langle HV|\hat{\rho}|HV\rangle +
\langle VH|\hat{\rho}|VH\rangle +
\langle VV|\hat{\rho}|VV\rangle \right).
\end{eqnarray}
Thus we obtain the final formula for the tomographic reconstruction
of the density matrices of our states:
\begin{equation}
\hat{\rho} =\left(\sum_{\nu=1}^{16}\hat{M}_{\nu} n_{\nu} \right)/
\left(\sum_{\nu=1}^{4} n_{\nu}\right).
\label{leopold}
\end{equation}

As an example, the following set of 16 counts were taken for
the purpose of tomographically determining the density matrix
for an ensemble of qubits all prepared in a specific quantum
state:
$ n_{1} = 34749,
n_{2} = 324,
n_{3} = 35805,
n_{4} = 444,
n_{5} = 16324,
n_{6} = 17521,
n_{7} = 13441,
n_{8} = 16901,
n_{9} = 17932,
n_{10} = 32028,
n_{11} = 15132,
n_{12} = 17238,
n_{13} = 13171,
n_{14} = 17170,
n_{15} = 16722,
n_{16} = 33586$
Applying eq.(\ref{leopold}) we find
\begin{equation}
\hat{\rho} =\left(
\begin{array}{cccc}
  0.4872            & -0.0042 + i 0.0114 & -0.0098 - i 0.0178 & 0.5192
+ i 0.0380\\
-0.0042 - i 0.0114 & 0.0045             &  0.0271 - i 0.0146 &
-0.0648 - i 0.0076\\
-0.0098 + i 0.0178 & 0.0271 + i 0.0146  &  0.0062            &
-0.0695 + i 0.0134 \\
  0.5192 - i 0.0380 & -0.0648 +i 0.0076  & -0.0695 - i 0.0134 & 0.5020
\end{array}
\right)
\end{equation}
This matrix is shown graphically in fig.3a.

Note that, by construction, the density matrix is normalized, i.e.
${\rm Tr}\{\hat{\rho}\} = 1$ and Hermitian, i.e.
$\hat{\rho}^{\dagger} = \hat{\rho}$ .
However, when one calculates the eigenvalues of this measured density
matrix, one finds the values  1.02155, 0.0681238, -0.065274 and 
-0.024396; and also, ${\rm Tr}\{\hat{\rho}^{2}\} = 1.053$ .
Density matrices for all physical states must have the property of
positive semi-definiteness, which (in conjunction with the
normalization and Hermiticity properties) imply that all of the
eigenvalues must lie in the interval $[0,1]$, their sum being 1; this 
in turn implies that $0\le {\rm Tr}\{\hat{\rho}^{2}\} \le 1$.
Clearly the density matrix reconstructed above by linear tomography
violates these condition.  From our experience of tomographic measurements
of various mixed and entangled states prepared experimentally, this
seems to happen roughly 75$\%$ of the time for low entropy, highly 
entangled states; it seems to have a higher probability of producing
the correct result for states of higher entropy, but the cautious
experimenter should check every time.  The obvious culprit for
this problem is experimental inaccuracies and statistical fluctuations of
coincidence counts, which mean that the actual numbers of counts
recorded in a real experiment differ from those that can be calculated
by eq.(\ref{hornby}).  Thus the linear reconstruction is of limited
value for states of low entropy (which are of most experimental 
interest because of their application to quantum information 
technology); however as we shall see, the linear approach does provide a useful
starting point for the numerical optimization approach to density 
matrix estimation which we will discuss in the next section.

\section{Maximum Likelihood Estimation}
\setcounter{equation}{0}
As mentioned in Section III, the tomographic measurement of density matrices
can produce results which violate important basic properties such as
positivity.  To avoid this problem, the maximum likelihood estimation
of density matrices may be employed.  Here we describe a simple
realization of this technique.

\subsection{Basic approach}
Our approach to the maximum likelihood estimation
of the density matrix is as follows:

\noindent
(i) Generate a formula for an explicitly ``physical''
density matrix, i.e. a matrix which has the three important
properties of normalization, Hermiticity and positivity. This
matrix will be a function of 16 real variables (denoted $\{ t_{1},t_{2},
\ldots t_{16} \}$).  We will
denote the matrix as $\hat{\rho}_{p}(t_{1},t_{2}, \ldots t_{16})$.

\noindent
(ii) Introduce a
``likelihood function'' which quantifies how good
the density matrix $\hat{\rho}_{p}(t_{1},t_{2}, \ldots t_{16})$
is in relation to the experimental data.  This likelihood
function is a function of the 16 real parameters $t_{\nu}$
and of the 16 experimental data $n_{\nu}$.  We will denote
this function as ${\cal L} (t_{1},t_{2}, \ldots t_{16}; n_{1},n_{2}, \ldots
n_{16} ) $.

\noindent
(iii) Using standard numerical optimization techniques, find
the optimum set of variables $\{t^{(opt)}_{1},t^{(opt)}_{2}, \ldots
t^{(opt)}_{16}\}$
for which the function ${\cal L} (t_{1},t_{2}, \ldots t_{16};
n_{1},n_{2}, \ldots
n_{16} ) $ has its maximum value.  The best estimate for the density
matrix is then $\hat{\rho} (t^{(opt)}_{1},t^{(opt)}_{2}, \ldots
t^{(opt)}_{16}) $.

The details of how these three steps can be carried out are described
in the next three sub-sections.

\subsection{Physical Density Matrices}
The property of non-negative definiteness for any matrix
$\hat{{\cal G}}$ is written mathematically as
\begin{equation}
\langle \psi | \hat{{\cal G}} |\psi\rangle \ge 0 \;\;\;\; \forall
|\psi\rangle.
\label{nnd}
\end{equation}
{\em Any} matrix that can be written in the form $\hat{{\cal G}} =
\hat{T}^{\dagger}\hat{T} $ must be non-negative definite.  To see that
this is the case, substitute into eq.(\ref{nnd}):
\begin{equation}
\langle \psi | \hat{T}^{\dagger}\hat{T} |\psi\rangle =
  \langle \psi^{\prime} | \psi^{\prime}\rangle \ge 0 ,
  \end{equation}
where we have defined $| \psi^{\prime}\rangle = \hat{T} |\psi\rangle$.
Furthermore $ (\hat{T}^{\dagger}\hat{T})^{\dagger} =
\hat{T}^{\dagger}(\hat{T}^{\dagger})^{\dagger}
=\hat{T}^{\dagger}\hat{T}$, i.e. $\hat{{\cal G}} =
\hat{T}^{\dagger}\hat{T} $ must be Hermitian.  To ensure
normalization, one can simply divide by the trace: thus
the matrix $\hat{g}$ given by the formula
\begin{equation}
\hat{g} =
\hat{T}^{\dagger}\hat{T}/
{\rm Tr}\{\hat{T}^{\dagger}\hat{T}\}
\label{jamesbond}
\end{equation}
has all three of the mathematical properties which we require for
density matrices.

For the two qubit system, we have a $4\times 4$ density matrix with
15 independent real parameters. Since it will be useful to be able
to invert relation (\ref{jamesbond}), it is convenient to
choose a tri-diagonal form for $\hat{T}$:
\begin{equation}
\hat{T}(t) = \left(\begin{array}{cccc}
t_{1}                 & 0                 & 0            & 0\\
t_{5}+i t_{6}         & t_{2}             & 0            & 0\\
t_{11}+i t_{12}         & t_{7}+i t_{8}     & t_{3}        & 0\\
t_{15}+i t_{16}         & t_{13}+i t_{14}     &t_{9}+i t_{10} & t_{4}
\end{array}\right),
\end{equation}

Thus the explicitly ``physical'' density matrix $\hat{\rho}_{p}$
is given by the formula
\begin{equation}
\hat{\rho}_{p}(t) =
\hat{T}^{\dagger}(t)\hat{T}(t)/
{\rm Tr}\{\hat{T}^{\dagger}(t)\hat{T}(t)\}.
\label{pussygalore}
\end{equation}

For future reference, the inverse relationship, by which the elements of
$\hat{T}$ can be expressed in terms of the elements of $\hat{\rho}$,
is as follows:
\begin{equation}
\hat{T}=\left(\begin{array}{cccc}
\sqrt{\frac{\Delta}{{\cal M}^{(1)}_{11}}}  &  0  &  0  &  0  \\
&&&\\
\frac{{\cal M}^{(1)}_{12}}{\sqrt{{\cal M}^{(1)}_{11}{\cal
M}^{(2)}_{11,22}}} & \sqrt{\frac{{\cal M}^{(1)}_{11}}{{\cal
M}^{(2)}_{11,22}}} &  0  & 0 \\&&&\\
\frac{{\cal M}^{(2)}_{12,23}}{\sqrt{\rho_{44}}\sqrt{{\cal
M}^{(2)}_{11,22}}} &
\frac{{\cal M}^{(2)}_{11,23}}{\sqrt{\rho_{44}}\sqrt{{\cal
M}^{(2)}_{11,22}}} &
\sqrt{\frac{{\cal M}^{(2)}_{11,22}}{\rho_{44}}} &
0 \\&&&\\
\frac{\rho_{41}}{\sqrt{\rho_{44}}} &
\frac{\rho_{42}}{\sqrt{\rho_{44}}} &
\frac{\rho_{43}}{\sqrt{\rho_{44}}} &
\sqrt{\rho_{44}}
\end{array}\right).
\label{invrel}
\end{equation}
Here we have used the notation $\Delta = {\rm Det}(\hat{\rho})$;
${\cal M}^{(1)}_{ij}$ is the first minor of $\hat{\rho}$, i.e. the
determinant of the $3\times 3$ matrix formed by deleting the
$i$-th row and $j$-th column of $\hat{\rho}$; ${\cal M}^{(2)}_{ij,kl}$ is
the second minor of $\hat{\rho}$, i.e. the
determinant of the $2\times 2$ matrix formed by deleting the
$i$-th and $k$-th rows and $j$-th and $l$-th columns of $\hat{\rho}$
($i\neq k$ and $j\neq l$).

\subsection{The Likelihood Function}
The measurement data consists of a set of 16 coincidence counts
$n_{\nu}\;(\nu=1, 2, \ldots 16)$
whose expected value is 
$\bar{n}_{\nu}={\cal N} \langle\psi_{\nu}|\hat{\rho}|\psi_{\nu}\rangle$. 
Let us assume
that the noise on these coincidence measurements has a Gaussian
probability distribution.  Thus the probability of obtaining a set
of 16 counts $\{n_{1},n_{2}, \ldots N_{16} \}$ is
\begin{equation}
P\left(n_{1},n_{2}, \ldots n_{16}\right) = \frac{1}{{\em Norm}}
\prod_{\nu=1}^{16} \exp\left[ - \frac{(n_{\nu}-\bar{n}_{\nu})^{2}}{2
\sigma_{\nu}^{2}}\right],
\end{equation}
where $\sigma_{\nu}$ is the standard deviation for $\nu$-th
coincidence measurement (given approximately by $\sqrt{\bar{n}_{\nu}}$)
and ${\em Norm}$ is the normalization constant.  For
our physical density matrix $\hat{\rho}_{p}$ the number
of counts expected for the $\nu$-th measurement is
\begin{equation}
\bar{n}_{\nu}\left(t_{1},t_{2}, \ldots t_{16}\right) =
{\cal N} \langle\psi_{\nu}|
\hat{\rho}_{p}\left(t_{1},t_{2}, \ldots t_{16}\right)
|\psi_{\nu}\rangle .
\end{equation}
Thus the likelihood that the matrix $\hat{\rho}_{p}\left(t_{1},t_{2},
\ldots t_{16}\right)$ could
produce the measured data $\{n_{1}, n_{2}, \ldots n_{16} \}$ is
\begin{equation}
P\left(t_{1},t_{2}, \ldots t_{16}\right) = \frac{1}{{\em Norm}}
\prod_{\nu=1}^{16} 
\exp\left[
- \frac{
({\cal N} \langle\psi_{\nu}|\hat{\rho}_{p}\left(t_{1},t_{2}, \ldots t_{16}\right)
|\psi_{\nu}\rangle - n_{\nu})^{2}}
{2 {\cal N}
\langle\psi_{\nu}|\hat{\rho}_{p}\left(t_{1},t_{2}, \ldots t_{16}\right)
|\psi_{\nu}\rangle - n_{\nu})}\right],
\end{equation}
where ${\cal N} = \sum_{\nu=1}^{4} N_{\nu} $.

Rather than find maximum value of $P\left(t_{1},t_{2}, \ldots
t_{16}\right)$ it simplifies things somewhat to find the maximum
of its logarithm (which is mathematically equivalent)
\footnote{Note that here we neglect the dependence of the 
normalization constant on $t_{1},t_{2}, \ldots
t_{16}$, which only weakly effects solution for
the most likely state}.  Thus
the optimization problem reduces to finding the {\em minimum}
of the following function:
\begin{equation}
{\cal L} \left(t_{1},t_{2}, \ldots t_{16}\right) =
\sum_{\nu=1}^{16}
\frac{
({\cal N} \langle\psi_{\nu}|\hat{\rho}_{p}\left(t_{1},t_{2}, \ldots t_{16}\right)
|\psi_{\nu}\rangle - n_{\nu})^{2}}
{2 {\cal N}
\langle\psi_{\nu}|\hat{\rho}_{p}\left(t_{1},t_{2}, \ldots t_{16}\right)
|\psi_{\nu}\rangle - n_{\nu})}.
\end{equation}
This is the ``likelihood'' function which we employed in our
numerical optimization routine.

\subsection{Numerical Optimization}
We used the {\tt Mathematica 4.0} routine {\tt FindMinimum}
which executes a multidimensional Powell direction set algorithm
(see ref.\cite{numrep} for a description of this algorithm).
To execute this routine, one requires an initial estimate for the
values of $t_{1},t_{2}, \ldots t_{16}$.  For this, we used the
tomographic estimate of the density matrix in the inverse
relation (\ref{invrel}), allowing us to determine a set of
values for $t_{1},t_{2}, \ldots t_{16}$.  Since the tomographic
density matrix may not be non-negative definite, the
values of the $t_{\nu}$'s deduced in this manner are not
necessarily real.  Thus for our initial guess we used the
real parts of the $t_{\nu}$'s deduced from the tomographic density
matrix.

For the example given in Section 2, the maximum likelihood
estimate is
\begin{equation}
\hat{\rho} =\left(
\begin{array}{cccc}
0.5069             & -0.0239 + i 0.0106 & -0.0412 - i  0.0221 & 0.4833 + i 0.0329 \\
-0.0239 - i 0.0106 & 0.0048             & 0.0023 + i  0.0019 & -0.0296 - i 0.0077\\
-0.0412 + i 0.0221 & 0.0023 - i 0.0019 & 0.0045              & -0.0425 + i 0.0192\\
0.4833 - i 0.0329 & -0.0296 + i 0.0077 & -0.0425 - i 0.0192 & 0.4839
\end{array}
\right).
\end{equation}
This matrix is illustrated in Fig.2b.  In this case, the matrix has 
eigenvalues 0.986022, 0.0139777, 0 and 0; and
${\rm Tr}\{\hat{\rho}^{2}\} = 0.972435$, indicating, while
the linear reconstruction gave a non-physical density matrix,
the maximum-likelihood reconstruction gives a legitimate density matrix.
\section{Error Analysis}
\setcounter{equation}{0}

In this section we present an analysis of the errors inherent in the
tomographic scheme described in Section III.  Two sources of errors
are found to be important: the shot noise error in the measured
coincidence counts $n_{\nu}$ and the uncertainty in the settings of
the angles of the waveplates used to make the tomographic projection
states.  We will analyze these two sources separately.

In addition to determining the density matrix of a pair of qubits,
one is often also interested in quantities derived from the
density matrix, such as the entropy or the entanglement of formation.
For completeness, we will also derive the errors in some of
these quantities.

\subsection{Errors due to Count Statistics}
 From eq.(\ref{leopold}) we see that the density matrix is specified
by a set of sixteen parameters $s_{\nu}$ defined by
\begin{equation}
s_{\nu}=n_{\nu}/{\cal N},
\label{george}
\end{equation}
where $n_{\nu}$ are the measured coincidence counts
and ${\cal N} =\sum_{\nu=1}^{4} n_{\nu}$.
We can determine the errors in $s_{\nu}$ using the
following formula \cite{melissinos}
\begin{equation}
\overline{\delta s_{\nu}\delta s_{\mu}}=\sum_{\lambda,\kappa = 1}^{16}
\left(\frac{\partial s_{\nu}}{\partial n_{\lambda}} \right)
\left(\frac{\partial s_{\mu}}{\partial n_{\kappa}} \right)
\overline{\delta n_{\lambda}\delta n_{\kappa}},
\label{heavyside}
\end{equation}
where the over-bar denotes the ensemble average of the random
uncertainties $\delta s_{\nu}$ and $\delta n_{\lambda}$.
The measured coincidence counts $n_{\lambda}$ are
statistically independent Poissonian random variables,
which implies the following relation:
\begin{equation}
\overline{\delta n_{\lambda}\delta n_{\kappa}} = n_{\lambda}
\delta_{\lambda,\kappa},
\label{poisson}
\end{equation}
where $\delta_{\lambda,\kappa}$ is the Kronecker delta.

Taking the derivative of eq.(\ref{george}), we find that
\begin{equation}
\frac{\partial s_{\mu}}{\partial n_{\nu}} =
\frac{1}{{\cal N}}\delta_{\mu\nu}-\frac{n_{\mu}}{{\cal N}^{2}}D_{\nu} ,
\label{hornet}
\end{equation}
where
\begin{equation}
D_{\nu}  = \sum_{\lambda=1}^{4} \delta_{\lambda,\nu}
= \left\{
\begin{array}{ll}
1 & \mbox{if}\,\,\, 1\le\nu\le 4\\
&\\
0 & \mbox{if}\,\,\, 5\le\nu\le 16 .
\end{array} \right.
\label{stokes}
\end{equation}
Substituting from eq.(\ref{hornet}) into eq.(\ref{heavyside})
and using eq.(\ref{poisson}), we obtain the result
\begin{equation}
\overline{\delta s_{\nu}\delta s_{\mu}}=
\frac{n_{\mu}}{{\cal N}^{2}}\delta_{\nu,\mu}
+\frac{n_{\nu}n_{\mu}}{{\cal N}^{3}} (1-D_{\mu}-D_{\nu}).
\label{maxwell}
\end{equation}
In most experimental circumstances ${\cal N}\gg 1$, and so the
second term in eq.(\ref{maxwell}) is negligibly small in comparison
to the first.  We shall therefore ignore it, and use the approximate
expression in the subsequent discussion;
\begin{equation}
\overline{\delta s_{\nu}\delta s_{\mu}}\approx
\frac{n_{\mu}}{{\cal N}^{2}}\delta_{\nu,\mu} \equiv
\frac{s_{\mu}}{{\cal N}}\delta_{\nu,\mu}  .
\label{cavendish}
\end{equation}

\subsection{Errors due to Angular Settings Uncertainties}
Using the formula (\ref{weetabix}) for the parameters
$s_{\nu}$ we can find the dependence of the measured density
matrix on errors in the tomographic states.  The derivative
of $s_{\nu}$ with respect to some generic wave-plate settings angle
$\theta$ is
\begin{equation}
\frac{\partial s_{\nu}}{\partial\theta} =
\left\{ \frac{\partial}{\partial\theta} \langle \psi_{\nu} | \right\}
\hat{\rho} |\psi_{\nu}\rangle
+
\langle \psi_{\nu}|
\hat{\rho} \left\{ \frac{\partial}{\partial\theta}|\psi_{\nu}\rangle \right\},
\label{newton}
\end{equation}
where $|\psi_{\nu}\rangle$ is the ket of the $\nu$-th projection
state [see eq.(\ref{state})].   Substituting from eq.(\ref{tumblety})
we find
\begin{equation}
\frac{\partial s_{\nu}}{\partial\theta} =
\sum_{\mu=1}^{16} s_{\mu} \left[
\left\{\frac{\partial}{\partial\theta} \langle \psi_{\nu} | \right\}
\hat{M_{\mu}} |\psi_{\nu}\rangle
+
\langle \psi_{\nu}|
\hat{M_{\mu}} \left\{
\frac{\partial}{\partial\theta}|\psi_{\nu}\rangle \right\}
\right] .
\end{equation}

For convenience, we shall label the four waveplate angles
$\{h_{1,\nu},q_{1,\nu},h_{2,\nu},q_{2,\nu}\}$ which
specify the $\nu$-th state by
$\{\theta_{\nu,1},\theta_{\nu,2},\theta_{\nu,3},\theta_{\nu,4}\}$
respectively.  Clearly the $\mu$-th state does not depend on
any of the $\nu$-th set of angles.  Thus we obtain the following
expression for the derivatives of $s_{\nu}$ with respect to
waveplate settings:
\begin{equation}
\frac{\partial s_{\nu}}{\partial\theta_{\lambda,i}} =
\delta_{\nu,\lambda}
\sum_{\mu=1}^{16} s_{\mu} f^{(i)}_{\nu,\mu},
\label{hooke}
\end{equation}
where
\begin{equation}
f^{(i)}_{\nu,\mu} =
\left\{\frac{\partial}{\partial\theta_{\nu,i}} \langle \psi_{\nu} | \right\}
\hat{M_{\mu}} |\psi_{\nu}\rangle
+
\langle \psi_{\nu}|
\hat{M_{\mu}} \left\{
\frac{\partial}{\partial\theta_{\nu,i}}|\psi_{\nu}\rangle \right\}.
\end{equation}
The 1024 quantities $f^{(i)}_{\nu,\mu}$ can be determined by taking the
derivatives of the functional forms of the tomographic states given by
eqs.(\ref{yamato}) and (\ref{state}), and evaluating those derivatives
at the appropriate values of the arguments (see Table 1).

The errors in the angles are assumed to be uncorrelated, as would be
the case if each wave-plate were adjusted for each of the 16
measurements.  In reality, for qubit experiments, only one or two of the
four waveplates are
adjusted between one measurement and the next.  However the assumption
of uncorrelated angular errors greatly simplifies the calculation  (which is,
after all, only an {\em estimate} of the errors), and seems to produce
reasonable figures for our error bars \footnote{In other experimental
circumstances, such as the measurement of the joint state of two
spin $1/2$ particles, the tomography would be realized by performing
unitary operations on the spins prior to measurement.  In this case,
an assumption analogous to ours will be wholly justified.}
Thus with the following assumption
\begin {equation}
\overline{\delta \theta_{\nu,i}\delta \theta_{\mu,j}}=
\delta_{\nu,\mu}\delta_{i,j} (\Delta\theta)^{2},
\end{equation}
(where $\Delta\theta$ is the RMS uncertainty in the setting
of the waveplate, with an estimated value of $0.25^{o}$ for our apparatus)
we obtain the following expression for the errors in
$s_{\nu}$ due to angular settings:
\begin{equation}
\overline{\delta s_{\nu}\delta s_{\mu}}=
\delta_{\nu,\mu}\sum_{i=1}^{4}\sum_{\epsilon,\lambda=1}^{16}
f^{(i)}_{\nu,\epsilon}f^{(i)}_{\nu,\lambda}
s_{\epsilon}s_{\lambda}
\label{rayleigh}
\end{equation}

Combining eqs.(\ref{rayleigh}) and (\ref{cavendish}) we obtain
the following formula for the total error in the quantities
$s_{\nu}$:
\begin{equation}
\overline{\delta s_{\nu}\delta s_{\mu}}=\delta_{\nu,\mu} \Lambda_{\nu}
\end{equation}
where
\begin{equation}
\Lambda_{\nu} = \left[\frac{s_{\nu}}{{\cal N}}+
\sum_{i=1}^{4}\sum_{\epsilon,\lambda=1}^{16}
f^{(i)}_{\nu,\epsilon}f^{(i)}_{\nu,\lambda}
s_{\epsilon}s_{\lambda}\right].
\label{frobisher}
\end{equation}
These sixteen quantities can be calculated using the parameters
$s_{\nu}$ and the constants $f^{(i)}_{\nu,\epsilon}$.  Note that
the same result can be obtained by assuming {\em a priori} that
the errors in the $s_{\nu}$ are all uncorrelated, with
$\Lambda_{\nu}= \delta s_{\nu}^{2}$; the more 
rigorous treatment given here is however necessary to demonstrate
this fact.  For a typical number of counts, say $\cal{N} = 10000$
it is found that the contribution of errors from the two causes
is roughly comparable; for larger numbers of counts, the angular
settings will become the dominant source of error.

Based on these results, the errors in the values of the various elements of
the density matrix estimated by the linear tomographic technique
described in Section 3 are as follows:
\begin{eqnarray}
\left(\Delta \rho_{i,j}\right)^{2}&=&
\sum_{\nu,\mu=1}^{16} \frac{\partial\rho_{i,j}}{\partial s_{\nu}}
\frac{\partial\rho_{i,j}}{\partial s_{\mu}}
\overline{\delta s_{\nu}\delta s_{\mu}} \nonumber \\
&=&
\sum_{\nu=1}^{16} \left(M_{\nu(i,j)}\right)^{2}
\Lambda_{\nu}
\label{hawkins}
\end{eqnarray}
where $M_{\nu(i,j)}$ is the $i,j$ element of the matrix $\hat{M}_{\nu}$.

A convenient way in which to estimate errors for a maximum
likelihood tomographic technique (rather than a linear tomographic
technique) is to employ
the above formulae, with the slight modification that the
parameter $s_{\nu}$ should be recalculated from eq.(\ref{weetabix})
using the estimated density matrix $\hat{\rho}_{est}$.  This does
not take into account errors inherent in the maximum likelihood
technique itself.

\subsection{Errors in Quantities Derived from the Density Matrix}
When calculating the propagation of errors, it is actually more convenient
to use the errors in the $s_{\nu}$ parameters, (given by
eq.(\ref{frobisher}), rather than the errors in the elements of
density matrix itself (which have non-negligible correlations).

\subsubsection{von Neumann Entropy}
The von Neumann entropy is an important measure of the 
purity of a quantum state $\hat{\rho}$.  It is
defined by \cite{mikeandike}
\begin{eqnarray}
\cal{S} &=& -{\rm Tr}\left\{
\hat{\rho}\log_{2}\left(\hat{\rho}\right)\right\} \nonumber \\
& = & -\sum_{a=1}^{4}p_{a} \log_{2} p_{a},
\label{entdef}
\end{eqnarray}
where 
$p_{a}$ is an eigenvalue of $\hat{\rho}$, i.e.
\begin{equation}
\hat{\rho}|\phi_{a}\rangle = p_{a}|\phi_{a}\rangle,
\end{equation}
$|\phi_{a}\rangle$ being the $a$-th eigenstate ($a$= 1,\ldots 4).
The error in this quantity is given by
\begin{equation}
\left(\Delta \cal{S}\right)^{2} =
\sum_{\nu=1}^{16}\left(\frac{\partial \cal{S}}{\partial
s_{\nu}}\right)^{2}
\Lambda_{\nu}.
\end{equation}
Applying the chain rule, we find
\begin{equation}
\left(\frac{\partial \cal{S}}{\partial
s_{\nu}}\right) = \sum_{a=1}^{4}
\left(\frac{\partial p_{a}}{\partial
s_{\nu}}\right)
\left(\frac{\partial \cal{S}}{\partial
p_{a}}\right).
\label{wasp}
\end{equation}
The partial differential 
of a eigenvalue can be easily found by perturbation theory.
As is well known (e.g. \cite{schiff}) the change in the
eigenvalue $\lambda_{a}$ of a matrix $\hat{W}$ due to a perturbation
in the matrix $\hat{\delta W}$ is
\begin{equation}
\delta\lambda_{a} =
\langle \phi_{a} | \hat{\delta W} |\phi_{a}\rangle ,
\end{equation}
where $|\phi_{a}\rangle$ is the eigenvector of $\hat{W}$
corresponding to the eigenvalue $\lambda_{a}$.  Thus the
derivative of $\lambda_{a}$ with respect to some variable
$x$ is given by
\begin{equation}
\frac{\partial \lambda_{a}}{\partial x}
=\langle \phi_{a} | 
\frac{\partial \hat{W}}{\partial x} |\phi_{a}\rangle .
\end{equation}
Since $\hat{\rho} = \sum_{\nu=1}^{16} \hat{M}_{\nu}
s_{\nu} $, we find that 
\begin{equation}
\frac{\partial p_{a}}{\partial
s_{\nu}} = \langle \phi_{a} | \hat{M}_{\nu}|\phi_{a}\rangle
\end{equation}
and so, taking the derivative of eq.(\ref{entdef}),
eq.(\ref{wasp}) becomes
\begin{equation}
\left(\frac{\partial \cal{S}}{\partial
s_{\nu}}\right) = -\sum_{a=1}^{4} \langle \phi_{a} | \hat{M}_{\nu}|\phi_{a}\rangle
\frac{\left[1+\ln p_{a}\right]}{\ln 2}.
\end{equation}
Hence
\begin{equation}
\left(\Delta \cal{S}\right)^{2} =
\sum_{\nu=1}^{16}
\left(
\sum_{a=1}^{4} \langle \phi_{a} | \hat{M}_{\nu}|\phi_{a}\rangle
\frac{\left[1+\ln p_{a}\right]}{\ln 2}
\right)^{2}
\Lambda_{\nu}.
\end{equation}
For the experimental example given above, ${\cal S} = 0.106 \pm 0.049 $.

\subsubsection{Linear Entropy}
The ``linear entropy'' is used to quantify the degree of mixture of a quantum state
in an analytically convenient form, although unlike the von Neumann entropy it has 
no direct information theoretic implications.  In a normalized form
(defined so that its value lies between zero and one), the linear entropy for a two
qubit system is defined by:
\begin{eqnarray}
\cal{P} &=& \frac{4}{3} \left(1-{\rm Tr}\left\{
\hat{\rho}^{2}\right\}\right) \nonumber \\
&=& \frac{4}{3} \left(1-
\sum_{a=1}^{4} p^{2}_{a} \right) .
\end{eqnarray}
To calculate the error in this quantity, we need the
following partial derivative:
\begin{eqnarray}
\frac{\partial \cal{P}}{\partial s_{\nu}} & =&
-\frac{8}{3} \sum_{a=1}^{4} p_{a} \frac{\partial p_{a}}
{\partial s_{\nu}} \nonumber \\
& =&-\frac{8}{3} \sum_{a=1}^{4} p_{a}
\langle \phi_{a} | \hat{M}_{\nu}|\phi_{a}\rangle \nonumber \\
& =&-\frac{8}{3} {\rm Tr}\left\{
\hat{\rho} \hat{M}_{\nu} \right\} \nonumber \\
& =&-\frac{8}{3} \sum_{\mu=1}^{16} {\rm Tr} \left\{
\hat{M}_{\mu} \hat{M}_{\nu} \right\} s_{\mu}
\end{eqnarray}
Hence the error in the linear entropy is
\begin{eqnarray}
\left(\Delta \cal{P}\right)^{2} &= &
\sum_{\nu=1}^{16}\left(\frac{\partial \cal{P}}{\partial
s_{\nu}}\right)^{2}
\Lambda_{\nu}, \nonumber \\
&=&
\sum_{\nu}^{16}
\left(
\frac{8}{3} \sum_{\mu=1}^{16} {\rm Tr} \left\{
\hat{M}_{\mu} \hat{M}_{\nu} \right\} s_{\mu}
\right)^{2}
\Lambda_{\nu}.
\end{eqnarray}
For the example given in Sections III and IV, ${\cal P} = 0.037 \pm 0.026 $.

\subsubsection{Concurrence, Entanglement of Formation and Tangle}
The concurrence, entanglement of formation and tangle are
measures of the quantum-coherence properties of a mixed quantum state
\cite{wooters}.  
For two qubits \footnote{The analysis in this subsection applies to
the two qubit case only.  Measures of entanglement for mixed n-qubit
systems are a subject of on-going research: see, for example,
\cite{terhal} for a recent survey.}
, concurrence is defined as follows: consider the non-Hermitian
matrix $\hat{R} = \hat{\rho}\hat{\Sigma}\hat{\rho}^{{\rm T}}\hat{\Sigma}$
where the superscript ${\rm T}$ denotes transpose and the ``spin flip
matrix'' $\hat{\Sigma}$ is defined by:
\begin{equation}
\hat{\Sigma} = \left(
\begin{array}{cccc}
0&0&0&-1\\
0&0&1&0\\
0&1&0&0\\
-1&0&0&0
\end{array}
\right).
\end{equation}
Note that the definition of $\hat{\Sigma}$ depends on the
basis chosen; we have assumed here the ``computational
basis'' $\left\{|HH\rangle, |HV\rangle, |VH\rangle,
|VV\rangle\right\}$.
In what follows, it will be convenient to write
$\hat{R}$ in the following form:
\begin{equation}
\hat{R} = \frac{1}{2}\sum_{\mu,\nu =1}^{16} \hat{q}_{\mu,\nu}
s_{\mu}s_{\nu},
\end{equation}
where $\hat{q}_{\mu,\nu} =
\hat{M}_{\mu}\hat{\Sigma}\hat{M}_{\nu}^{{\rm T}}\hat{\Sigma}+
\hat{M}_{\nu}\hat{\Sigma}\hat{M}_{\mu}^{{\rm T}}\hat{\Sigma}$.
The left and right eigenstates and eigenvalues
of the matrix $\hat{R}$ we shall denote by $\langle\xi_{a}|$,
$|\zeta_{a}\rangle$ and $r_{a}$, respectively, i.e.:
\begin{eqnarray}
\langle\xi_{a}|\hat{R} &=& r_{a} \langle\xi_{a}| \nonumber \\
\hat{R}|\zeta_{a}\rangle &=& r_{a}|\zeta_{a}\rangle .
\end{eqnarray}
We shall assume that these eigenstates are normalized in
the usual manner for bi-orthogonal expansions, i.e.
$\langle\xi_{a}|\zeta_{b}\rangle = \delta_{a,b}$.  Further
we shall assume that the eigenvalues are numbered in decreasing
order, so that $r_{1}\ge r_{2}\ge r_{3}\ge r_{4}$.
The concurrence is then defined by the formula
\begin{eqnarray}
C &=& {\rm Max}\left\{0, \sqrt{r_{1}} - \sqrt{r_{2}} - \sqrt{r_{3}} -
\sqrt{r_{4}}\right\} \nonumber \\
&=& {\rm Max}\left\{ 0, \sum_{a=1}^{4}{\rm
sgn}\left(\frac{3}{2}-a\right) \sqrt{r_{a}}
\right\},
\end{eqnarray}
where ${\rm sgn}(x) = 1$ if $x>0$ and ${\rm sgn}(x) = -1$ if $x<0$.
The tangle is given by $T=C^{2}$ and the Entanglement of Formation
by
\begin{equation}
E = h\left(\frac{1+\sqrt{1-C^{2}}}{2}\right),
\end{equation}
where $h(x) =-x {\rm log}_{2}x-(1-x) {\rm log}_{2}(1-x) $.  Because
$h(x)$ is a monotonically increasing function, these three
quantities are to some extent equivalent measures of the
entanglement of a mixed state.

To calculate the errors in these rather complicated functions,
we must employ the perturbation theory for non-Hermitian
matrices (see Appendix C for more details).  
We need to evaluate the following partial derivative,
\begin{eqnarray}
\frac{\partial C}{\partial s_{\nu}} &=& \sum_{a=1}^{4}
{\rm sgn}\left(\frac{3}{2}-a\right) \frac{1}{2 \sqrt{r_{a}}}
\frac{\partial r_{a}}{\partial s_{\nu}} \nonumber\\
&=&\sum_{a=1}^{4}
{\rm sgn}\left(\frac{3}{2}-a\right) \frac{1}{2 \sqrt{r_{a}}}
\langle \xi_{a} |\frac{\partial \hat{R}}{\partial s_{\nu}}
|\zeta_{a}\rangle  \nonumber\\
&=&\sum_{a=1}^{4} \sum_{\mu=1}^{16}
{\rm sgn}\left(\frac{3}{2}-a\right) \frac{1}{2 \sqrt{r_{a}}}
\langle \xi_{a} |\hat{q}_{\mu,\nu} s_{\mu}|\zeta_{a}\rangle,
\end{eqnarray}
where the function ${\rm sgn}\left(x\right)$ is the sign of
the quantity $x$: it takes the value $1$ if $x > 0$ and
$-1$ if $x < 0$.  Thus ${\rm sgn}\left(3/2 - a\right)$
is equal to $+1$ if $a =1$ and $-1$ if $a =2, 3\,\, {\rm or} \,\,4$.
Hence the error in the concurrence is
\begin{eqnarray}
\left(\Delta C\right)^{2} &=& \sum_{\nu=1}^{16}
\left(
\frac{\partial C}{\partial s_{\nu}}
\right)^{2}
\Lambda_{\nu} \nonumber \\
&=& \sum_{\nu=1}^{16}
\left(
\sum_{a=1}^{4} \sum_{\mu=1}^{16}
{\rm sgn}\left(\frac{3}{2}-a\right) \frac{1}{2 \sqrt{r_{a}}}
\langle \xi_{a} |\hat{q}_{\mu,\nu} s_{\mu}|\zeta_{a}\rangle
\right)^{2}
\Lambda_{\nu}.
\end{eqnarray}
For our example the concurrence is
$0.963 \pm 0.018$.

Once we know the error in the concurrence, the errors in the tangle
and the entanglement of formation can be found straightforwardly:
\begin{eqnarray}
\Delta T &=& 2 C \Delta C \\
\Delta E &=& \frac{C}{\sqrt{1-C^{2}}} 
h^{\prime}\left(\frac{1+\sqrt{1-C^{2}}}{2}\right) \Delta C,
\end{eqnarray}
where $h^{\prime}(x)$ is the derivative of $h(x)$.
For our example the
the tangle is $0.928 \pm 0.034$ and
the entanglement of formation is $0.947 \pm 0.025$.

\section{Conclusions}
In conclusion, we have presented a technique for reconstructing
density matrices of qubit systems, including a full error analysis.
We have extended the latter through to calculation of quantities of
interest in quantum information, such as the entropy and concurrence.
Without loss of generality, we have used the example of polarization
qubits of entangled photons, but we stress that these techniques can
be adapted to any physical realization of qubits.

\section*{Acknowledgements}
The authors would like to thank
Joe Altepeter, Mauro d'Ariano, Zdenek Hradil, Kurt Jacobs,
Poul Jessen, Michael Neilsen, Mike Raymer, Sze Tan,
and Jaroslav \v{R}eh\'{a}\v{c}ek
for useful discussions and correspondence.  This
work was supported in part by the U.S. National Security Agency,
and Advanced Research and Development Activity (ARDA),
by the Los Alamos National Laboratory LDRD program
and by the Australian Research Council.


\section*{Appendix A: The $\hat{\Gamma}$-matrices}
\renewcommand{\theequation}{{\rm
A.\arabic{equation}}}\setcounter{equation}{0}

One possible set of $\hat{\Gamma}$-matrices are generators of
$SU(2)\otimes SU(2)$,
normalized so that the conditions given in eq.\ref{gammaconds} are
fulfilled.  These matrices are:
\begin{equation}
\begin{array}{cccc}
\hat{\Gamma}_{1} = \frac{1}{2} \left( \begin{array}{cccc}
0 & 1 & 0 & 0\\
1 & 0 & 0 & 0\\
0 & 0 & 0 & 1\\
0 & 0 & 1 & 0\\
\end{array}\right),&
\hat{\Gamma}_{2} = \frac{1}{2} \left( \begin{array}{cccc}
0 &-i & 0 & 0\\
i & 0 & 0 & 0\\
0 & 0 & 0 &-i\\
0 & 0 & i & 0\\
\end{array}\right),&
\hat{\Gamma}_{3} = \frac{1}{2} \left( \begin{array}{cccc}
1 & 0 & 0 & 0\\
0 &-1 & 0 & 0\\
0 & 0 & 1 & 0\\
0 & 0 & 0 &-1\\
\end{array}\right),&
\hat{\Gamma}_{4} = \frac{1}{2} \left( \begin{array}{cccc}
0 & 0 & 1 & 0\\
0 & 0 & 0 & 1\\
1 & 0 & 0 & 0\\
0 & 1 & 0 & 0\\
\end{array}\right),\\ \\
\hat{\Gamma}_{5} = \frac{1}{2} \left( \begin{array}{cccc}
0 & 0 & 0 & 1\\
0 & 0 & 1 & 0\\
0 & 1 & 0 & 0\\
1 & 0 & 0 & 0\\
\end{array}\right),&
\hat{\Gamma}_{6} = \frac{1}{2} \left( \begin{array}{cccc}
0 & 0 & 0 &-i\\
0 & 0 & i & 0\\
0 &-i & 0 & 0\\
i & 0 & 0 & 0\\
\end{array}\right),&
\hat{\Gamma}_{7} = \frac{1}{2} \left( \begin{array}{cccc}
0 & 0 & 1 & 0\\
0 & 0 & 0 &-1\\
1 & 0 & 0 & 0\\
0 &-1 & 0 & 0\\
\end{array}\right),&
\hat{\Gamma}_{8} = \frac{1}{2} \left( \begin{array}{cccc}
0 & 0 &-i & 0\\
0 & 0 & 0 &-i\\
i & 0 & 0 & 0\\
0 & i & 0 & 0\\
\end{array}\right),\\ \\
\hat{\Gamma}_{9} = \frac{1}{2} \left( \begin{array}{cccc}
0 & 0 & 0 &-i\\
0 & 0 &-i & 0\\
0 & i & 0 & 0\\
i & 0 & 0 & 0\\
\end{array}\right),&
\hat{\Gamma}_{10} = \frac{1}{2} \left( \begin{array}{cccc}
0 & 0 & 0 &-1\\
0 & 0 & 1 & 0\\
0 & 1 & 0 & 0\\
-1& 0 & 0 & 0\\
\end{array}\right),&
\hat{\Gamma}_{11} = \frac{1}{2} \left( \begin{array}{cccc}
0 & 0 &-i & 0\\
0 & 0 & 0 & i\\
i & 0 & 0 & 0\\
0 &-i & 0 & 0\\
\end{array}\right),&
\hat{\Gamma}_{12} = \frac{1}{2} \left( \begin{array}{cccc}
1 & 0 & 0 & 0\\
0 & 1 & 0 & 0\\
0 & 0 &-1 & 0\\
0 & 0 & 0 &-1\\
\end{array}\right),\\ \\
\hat{\Gamma}_{13} = \frac{1}{2} \left( \begin{array}{cccc}
0 & 1 & 0 & 0\\
1 & 0 & 0 & 0\\
0 & 0 & 0 &-1\\
0 & 0 &-1 & 0\\
\end{array}\right),&
\hat{\Gamma}_{14} = \frac{1}{2} \left( \begin{array}{cccc}
0 &-i & 0 & 0\\
i & 0 & 0 & 0\\
0 & 0 & 0 & i\\
0 & 0 &-i & 0\\
\end{array}\right),&
\hat{\Gamma}_{15} = \frac{1}{2} \left( \begin{array}{cccc}
1 & 0 & 0 & 0\\
0 &-1 & 0 & 0\\
0 & 0 &-1 & 0\\
0 & 0 & 0 & 1\\
\end{array}\right),&
\hat{\Gamma}_{16} = \frac{1}{2} \left( \begin{array}{cccc}
1 & 0 & 0 & 0\\
0 & 1 & 0 & 0\\
0 & 0 & 1 & 0\\
0 & 0 & 0 & 1\\
\end{array}\right).
\end{array}
\end{equation}
As noted in the text, this is only one possible choice for
these matrices, and the final results are independent of the
choice.

\section*{Appendix B: The $\hat{M}$-matrices and some of their
properties}
\renewcommand{\theequation}{{\rm
B.\arabic{equation}}}\setcounter{equation}{0}

The $\hat{M}$ matrices, defined by eq.(\ref{chapman}),
are as follows:

\begin{equation}
\begin{array}{cc}
\hat{M}_{1} = \frac{1}{2} \left( \begin{array}{cccc}
     2   & -(1-i) & -(1+i) & 1\\
  -(1+i) &    0   &    i   & 0\\
  -(1-i) &    -i  &    0   & 0\\
     1   &    0   &    0   & 0\\
\end{array}\right)
,&
\hat{M}_{2} = \frac{1}{2} \left( \begin{array}{cccc}
     0   & -(1-i) &    0   &     1    \\
  -(1+i) &    2   &    i   &  -(1+i)  \\
     0   &    -i  &    0   &     0    \\
     1   & -(1+i) &    0   &     0    \\
\end{array}\right)
,\\ \\
\hat{M}_{3} = \frac{1}{2} \left( \begin{array}{cccc}
     0   &    0   &    0   &    1   \\
     0   &    0   &    i   & -(1+i) \\
     0   &   -i   &    0   & -(1-i) \\
     1   & -(1-i) & -(1+i) &    2   \\
\end{array}\right)
,&
\hat{M}_{4} = \frac{1}{2} \left( \begin{array}{cccc}
     0   &    0   & -(1+i) &    1   \\
     0   &    0   &    i   &    0   \\
  -(1-i) &    -i  &    2   & -(1-i) \\
     1   &    0   & -(1+i) &    0   \\
\end{array}\right)
,\\ \\
\hat{M}_{5} = \frac{1}{2} \left( \begin{array}{cccc}
     0   &    0   &   2i   & -(1+i) \\
     0   &    0   &  (1-i) &    0   \\
    -2i  &  (1+i) &    0   &    0   \\
  -(1-i) &    0   &    0   &    0   \\
\end{array}\right)
,&
\hat{M}_{6} = \frac{1}{2} \left( \begin{array}{cccc}
     0   &    0   &    0   & -(1+i) \\
     0   &    0   &  (1-i) &   2i   \\
     0   &  (1+i) &    0   &    0   \\
  -(1-i) &  -2i   &    0   &    0   \\
\end{array}\right)
,\\ \\
\hat{M}_{7} = \frac{1}{2} \left( \begin{array}{cccc}
     0   &    0   &    0   & -(1+i) \\
     0   &    0   & -(1-i) &    2   \\
     0   & -(1+i) &    0   &    0   \\
  -(1-i) &    2   &    0   &    0   \\
\end{array}\right)
,&
\hat{M}_{8} = \frac{1}{2} \left( \begin{array}{cccc}
     0   &    0   &    2   & -(1+i) \\
     0   &    0   & -(1-i) &    0   \\
     2   & -(1+i) &    0   &    0   \\
  -(1-i) &    0   &    0   &    0   \\
\end{array}\right)
,\\ \\
\hat{M}_{9} = \left( \begin{array}{cccc}
     0   &    0   &    0   &    i   \\
     0   &    0   &   -i   &    0   \\
     0   &    i   &    0   &    0   \\
    -i   &    0   &    0   &    0   \\
\end{array}\right)
,&
\hat{M}_{10} =  \left( \begin{array}{cccc}
     0   &    0   &    0   &    1   \\
     0   &    0   &    1   &    0   \\
     0   &    1   &    0   &    0   \\
     1   &    0   &    0   &    0   \\
\end{array}\right)
,\\ \\
\hat{M}_{11} = \left( \begin{array}{cccc}
     0   &    0   &    0   &    i   \\
     0   &    0   &    i   &    0   \\
     0   &   -i   &    0   &    0   \\
    -i   &    0   &    0   &    0   \\
\end{array}\right)
,&
\hat{M}_{12} = \frac{1}{2} \left( \begin{array}{cccc}
     0   &    2   &    0   & -(1+i) \\
     2   &    0   & -(1+i) &    0   \\
     0   & -(1-i) &    0   &    0   \\
  -(1-i) &    0   &    0   &    0   \\
\end{array}\right)
,\\ \\
\hat{M}_{13} = \frac{1}{2} \left( \begin{array}{cccc}
     0   &    0   &    0   & -(1+i) \\
     0   &    0   & -(1+i) &    0   \\
     0   & -(1-i) &    0   &    2   \\
  -(1-i) &    0   &    2   &    0   \\
\end{array}\right)
,&
\hat{M}_{14} = \frac{1}{2} \left( \begin{array}{cccc}
     0   &    0   &    0   & -(1-i) \\
     0   &    0   & -(1-i) &    0   \\
     0   & -(1+i) &    0   &  -2i   \\
  -(1+i) &    0   &   2i   &    0   \\
\end{array}\right)
,\\ \\
\hat{M}_{15} = \frac{1}{2} \left( \begin{array}{cccc}
     0   &  -2i   &    0   & -(1-i) \\
    2i   &    0   &  (1-i) &    0   \\
     0   &  (1+i) &    0   &    0   \\
  -(1+i) &    0   &    0   &    0   \\
\end{array}\right)
,&
\hat{M}_{16} = \left( \begin{array}{cccc}
     0   &    0   &    0   &    1   \\
     0   &    0   &   -1   &    0   \\
     0   &   -1   &    0   &    0   \\
     1   &    0   &    0   &    0   \\
\end{array}\right).
\end{array}
\end{equation}

The form of these matrices is independent of the
chosen set of matrices $\left\{\hat{\Gamma}_{\nu}\right\}$
used to convert the density matrix into a column vector.
However the $\hat{M}_{\nu}$ matrices {\em do} depend on
the set of tomographic states $|\psi_{\nu}\rangle$.

There are some useful properties of these matrices
which we will now derive.
 From eq.(\ref{chapman}), we have
\begin{equation}
\langle\psi_{\mu}|\hat{M}_{\nu}|\psi_{\mu}\rangle =
\sum_{\lambda}\langle\psi_{\mu}|\hat{\Gamma}_{\lambda}|\psi_{\mu}\rangle
\left(B^{-1}\right)_{\lambda,\nu}.
\end{equation}
 From eq.(\ref{anderson}) we have
$\langle\psi_{\mu}|\hat{\Gamma}_{\lambda}|\psi_{\mu}\rangle=B_{\mu,\lambda}$,
thus we obtain the result
\begin{equation}
\langle\psi_{\mu}|\hat{M}_{\nu}|\psi_{\mu}\rangle = \delta_{\mu,\nu}.
\end{equation}

If we denote the basis set for the four-dimensional Hilbert space
by $\left\{ |i\rangle \,\, (i = 1,2,3,4)\right\}$, then
eq.(\ref{tumblety}) can be written as follows:
\begin{equation}
\langle i|\hat{\rho}| j \rangle=
\sum_{k,l}\sum_{\nu}\langle i|\hat{M}_{\nu}| j \rangle
\langle \psi_{\nu}| k \rangle
\langle l| \psi_{\nu} \rangle\langle k|\hat{\rho}| l \rangle .
\label{clarence}
\end{equation}
Since eq.(\ref{clarence}) is valid for arbitrary states $\hat{\rho}$,
we obtain the following relationship:
\begin{equation}
\sum_{\nu}\langle i|\hat{M}_{\nu}| j \rangle
\langle \psi_{\nu}| k \rangle
\langle l| \psi_{\nu} \rangle = \delta_{ik}\delta_{jl}.
\label{gull}
\end{equation}
Contracting eq.(\ref{gull}) over the indices $(i,j)$
we obtain:
\begin{equation}
\sum_{\nu}{\rm Tr}\left\{\hat{M}_{\nu}\right\}
| \psi_{\nu} \rangle\langle \psi_{\nu}| =
\hat{I},
\label{propone}
\end{equation}
where $\hat{I}$ is the identity operator for our four
dimensional Hilbert space.

A second relationship can be obtained by contracting eq.(\ref{gull}),
viz:
\begin{equation}
\sum_{\nu}\langle i|\hat{M}_{\nu}| j \rangle
  = \delta_{ij},
\end{equation}
or, in operator notation,
\begin{equation}
\sum_{\nu}\hat{M}_{\nu} = \hat{I}.
\end{equation}

\section*{Appendix C: Perturbation theory for non-Hermitian matrices}
\renewcommand{\theequation}{{\rm
C.\arabic{equation}}}\setcounter{equation}{0}

Whereas perturbation theory for Hermitian matrices is covered
in most quantum mechanics text-books, the case of non-Hermitian
matrices is less familiar, and so we will present it here.  The
problem is, given the eigenspectrum of a matrix $\hat{R}_{0}$
\cite{MF},
i.e.:
\begin{eqnarray}
\langle\xi_{a}|\hat{R}_{0} &=& r_{a} \langle\xi_{a}|  \\
\hat{R}_{0}|\zeta_{a}\rangle &=& r_{a}|\zeta_{a}\rangle ,
\end{eqnarray}
where
\begin{equation}
\langle\xi_{a}|\zeta_{b}\rangle = \delta_{a,b}
\label{biorth}
\end{equation}
we wish to find expressions for
the eigenvalues  $ r^{\prime}_{a}$ and eigenstates
$\langle  \xi^{\prime}_{a}|$ and $|\zeta^{\prime}_{a}\rangle$
of the perturbed matrix $\hat{R}^{\prime} = \hat{R}_{0}+\delta \hat{R}$.

We start with the standard assumption of perturbation theory,
i.e. that the perturbed quantities $ r^{\prime}_{a}$,
$\langle\xi^{\prime}_{a}|$ and $|\zeta^{\prime}_{a}\rangle$
can be expressed as power series of some parameter $\lambda$:
\begin{eqnarray}
r^{\prime}_{a} & = & r^{(0)}_{a} + \lambda r^{(1)}_{a}+\lambda^{2}
r^{(2)}_{a}+\ldots \\
|\zeta^{\prime}_{a}\rangle & = & |\zeta^{(0)}_{a}\rangle +
\lambda |\zeta^{(1)}_{a}\rangle+\lambda^{2} |\zeta^{(2)}_{a}\rangle+\ldots\\
\langle\xi^{\prime}_{a}| & = & \langle\xi^{(0)}_{a}| +
\lambda \langle\xi^{(1)}_{a}|+\lambda^{2} \langle\xi^{(2)}_{a}|+\ldots
\end{eqnarray}
Writing $\hat{R}^{\prime} = \hat{R}_{0}+ \lambda \delta \hat{R}$, and
comparing terms of equal powers of $\lambda$ in the eigen equations,
one obtains the following formulae:
\begin{eqnarray}
\hat{R}_{0}|\zeta^{(0)}_{a}\rangle &= &
r^{(0)}_{a}|\zeta^{(0)}_{a}\rangle \label{pestilence} \\
\langle\xi^{(0)}_{a}|\hat{R}_{0} &=&
r^{(0)}_{a} \langle\xi^{(0)}_{a}| \label{famine}\\
\left(\hat{R}_{0}-r^{(0)}_{a}\hat{I}\right)|\zeta^{(1)}_{a}\rangle
& =& -\left(\delta\hat{R}-r^{(1)}_{a}\right)|\zeta^{(0)}_{a}
\label{destruction}\rangle\\
\langle\xi^{(1)}_{a}|\left(\hat{R}_{0}-r^{(0)}_{a}\hat{I}\right)
&=&-\langle\xi^{(0)}_{a}|\left(\delta\hat{R}-r^{(1)}_{a}\right)
\label{death}.
\end{eqnarray}
Equations (\ref{pestilence}) and (\ref{famine}) imply that,
as might be expected,
\begin{eqnarray}
|\zeta^{(0)}_{a}\rangle & = &|\zeta_{a}\rangle\\
\langle\xi^{(0)}_{a}|& = &\langle\xi_{a}| \\
r^{(0)}_{a} &=&r_{a}
\end{eqnarray}
Taking the inner product of  eq.(\ref{destruction})
with $\langle\xi_{a}|$, and using the bi-orthogonal property
eq.(\ref{biorth}), we obtain
\begin{equation}
r^{(1)}_{a} = \langle\xi_{a}| \delta\hat{R} |\zeta_{a}\rangle.
\end{equation}
This implies that
\begin{eqnarray}
\delta r_{a} &\equiv& r^{\prime}_{a} - r_{a} \nonumber \\
&\approx & \langle\xi_{a}| \delta\hat{R} |\zeta_{a}\rangle.
\end{eqnarray}
Thus, dividing both sides by some differential increment
$\delta x $ and
taking the limit $\delta x \rightarrow 0$, we obtain
\begin{equation}
\frac{\partial r_{a}}{\partial x} =
\langle\xi_{a}| \frac{\partial \hat{R}}{\partial x} |\zeta_{a}\rangle.
\end{equation}

Using the completeness property of the eigenstates,
$ \sum_{b}|\zeta_{b}\rangle\langle\xi_{b}| = \hat{{\rm I}} $
and the identity $\hat{R}_{0} = \sum_{b} r_{b}
|\zeta_{b}\rangle\langle\xi_{b}|$
we obtain the following formula
\begin{equation}
\left(\hat{R}_{0}-r_{a}\hat{{\rm I}}\right)^{-1} =
\sum_{\stackrel{\scriptstyle b}{b \neq a}}
\frac{1}{r_{b}-r_{a}}|\zeta_{b}\rangle\langle\xi_{b}|.
\label{invrel2}
\end{equation}
Applying this to eq.(\ref{destruction}) we obtain
\begin{eqnarray}
|\delta \zeta^{(1)}_{a}\rangle &\equiv& |
\zeta^{\prime}_{a}\rangle -|\zeta_{a}\rangle \nonumber\\
&\approx&
-\sum_{\stackrel{\scriptstyle b}{b \neq a}}
\left(\frac{ \langle\xi_{b}| \delta\hat{R}
|\zeta_{a}\rangle}{r_{b}-r_{a}}\right) |\zeta_{b}\rangle .
\end{eqnarray}
Similarly, eqs.(\ref{death}) and (\ref{invrel2}) imply
\begin{eqnarray}
\langle\delta\xi_{a}| &\equiv&
\langle\delta\xi^{\prime}_{a}| -\langle\delta\xi_{a}| \nonumber\\
&\approx&
-\sum_{\stackrel{\scriptstyle b}{b \neq a}}
\left(\frac{\langle\xi_{a}| \delta\hat{R}
|\zeta_{b}\rangle}{r_{b}-r_{a}}\right) \langle\xi_{b}|.
\end{eqnarray}


\newpage

\begin{figure}[h]
\center{ \epsfig{figure=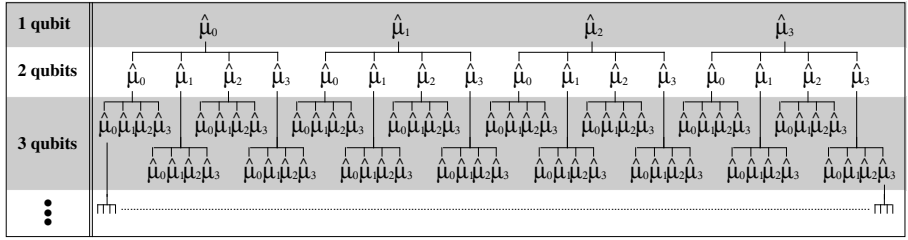,width=120mm}}
\caption{Tree diagram representing number and type of measurements
necessary for tomography.  For a single qubit, the measurements $\{
\hat{\mu}_{0}, \hat{\mu}_{1}, \hat{\mu}_{2}, \hat{\mu}_{3} \}$ suffice
to reconstruct the state, e.g., measurements of the horizontal,
vertical, diagonal and right-circular polarization components,
(H,V,D,R).  For two qubits, 16 double-coincidence measurements are
necessary ($\{ \hat{\mu}_{0} \hat{\mu}_{0}, \hat{\mu}_{0}
\hat{\mu}_{1} \ldots \hat{\mu}_{3} \hat{\mu}_{3} \}$), increasing to
64 three-coincidence measurements for three qubits ($\{ \hat{\mu}_{0}
\hat{\mu}_{0} \hat{\mu}_{0}, \hat{\mu}_{0} \hat{\mu}_{0} \hat{\mu}_{1}
\ldots \hat{\mu}_{3} \hat{\mu}_{3} \hat{\mu}_{3} \}$), and so on, as
shown.}
\label{fig1}
\end{figure}

\vspace{2cm}

\begin{figure}[h]
\center{ \epsfig{figure=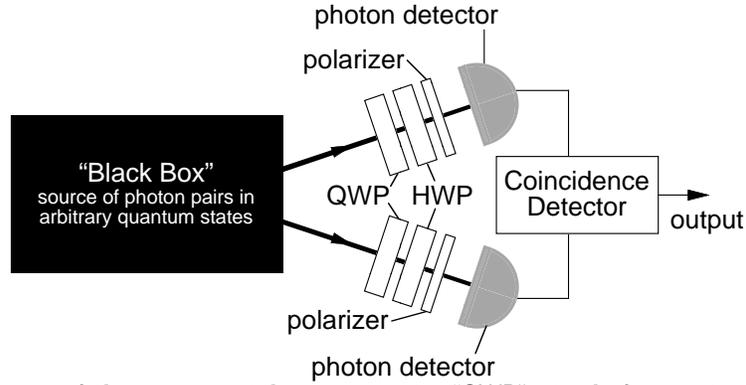,width=100mm}}
\caption{Schematic illustration of the experimental arrangement.
``{\sf QWP}'' stands for quarter waveplate, ``{\sf HWP}'' for half waveplate;
the angles of both pairs of wave plates can be set independently
giving the experimenter four degrees of freedom with which to set
the projection state.  In the experiment, the polarizers were realized
using polarizing prisms, arranged to transmit vertically
polarized light.}
\label{fig2}
\end{figure}

\vspace{2cm}

\begin{figure}[h]
\center{ \epsfig{figure=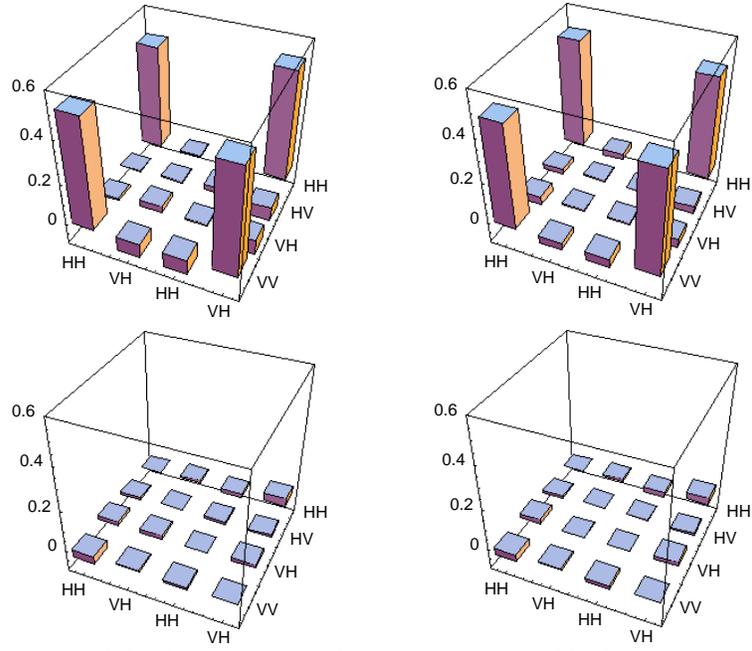,width=100mm}}
\caption{Graphical representation of the density matrix of
a state as estimated by linear tomography (left) and by maximum
likelihood tomography (right) from the experimental data given in the 
text.  The upper plot is the real part of
$\hat{\rho}$, the lower plot the imaginary part.}
\label{fig3}
\end{figure}

\newpage
\begin{center}
\begin{tabular}{|c|c|c|cccc|}\hline
$\nu$& Mode 1& Mode 2& $h_{1}$ & $ q_{1}$ & $h_{2}$ & $q_{2}$\\ \hline
1&$|{\rm H}\rangle$&$|{\rm H}\rangle$  & $45^{o}$   & $0$      & $45^{o}$   & $0$      \\
2&$|{\rm H}\rangle$&$|{\rm V}\rangle$  & $45^{o}$   & $0$      &   $0$      & $0$      \\
3&$|{\rm V}\rangle$&$|{\rm V}\rangle$  &   $0$      & $0$      &   $0$      & $0$      \\
4&$|{\rm V}\rangle$&$|{\rm H}\rangle$  &   $0$      & $0$      & $45^{o}$   & $0$      \\
5&$|{\rm R}\rangle$&$|{\rm H}\rangle$  & $22.5^{o}$ & $0$      & $45^{o}$   & $0$      \\
6&$|{\rm R}\rangle$&$|{\rm V}\rangle$ & $22.5^{o}$ & $0$      &   $0$      & $0$      \\
7&$|{\rm D}\rangle$&$|{\rm V}\rangle$ &$22.5^{o}$ & $45^{o}$ &   $0$      & $0$      \\
8&$|{\rm D}\rangle$&$|{\rm H}\rangle$  &$22.5^{o}$ & $45^{o}$ & $45^{o}$   & $0$      \\
9&$|{\rm D}\rangle$&$|{\rm R}\rangle$ &$22.5^{o}$ & $45^{o}$ & $22.5^{o}$ & $0$      \\
10&$|{\rm D}\rangle$&$|{\rm D}\rangle$  &$22.5^{o}$ & $45^{o}$ &$22.5^{o}$ & $45^{o}$ \\
11&$|{\rm R}\rangle$&$|{\rm D}\rangle$ & $22.5^{o}$ & $0$      &$22.5^{o}$ & $45^{o}$ \\
12&$|{\rm H}\rangle$&$|{\rm D}\rangle$  & $45^{o}$   & $0$      &$22.5^{o}$ & $45^{o}$ \\
13&$|{\rm V}\rangle$&$|{\rm D}\rangle$ &   $0$      & $0$      &$22.5^{o}$ & $45^{o}$ \\
14&$|{\rm V}\rangle$&$|{\rm L}\rangle$ &   $0$      & $0$      &$22.5^{o}$  & $90^{o}$ \\ 
15&$|{\rm H}\rangle$&$|{\rm L}\rangle$  & $45^{o}$   & $0$      &$22.5^{o}$  & $90^{o}$ \\
16&$|{\rm R}\rangle$&$|{\rm L}\rangle$ & $22.5^{o}$ & $0$      &$22.5^{o}$  & $90^{o}$ \\
\hline
\multicolumn{3}{c}{\bf Table 1}\\
\end{tabular}
\end{center}

TABLE 1: The tomographic analysis states used in our experiments.  The
number of coincidence counts measured in projections measurements
provide a set of 16 data that allow the density matrix of the state
of the two modes to be estimated.  We have used the notation
$|{\rm D}\rangle \equiv \left(|{\rm H}\rangle+|{\rm V}\rangle\right)
/\sqrt{2}$,
$|{\rm L}\rangle \equiv \left(|{\rm H}\rangle+i|{\rm V}\rangle
\right)/\sqrt{2}$ and
$|{\rm R}\rangle \equiv \left(|{\rm H}\rangle-i|{\rm V}\rangle \right)/\sqrt{2}$.
Note that, when the measurement are taken in the order given by the 
table, only one waveplate angle had to be changed between each 
measurement.

\vspace{3cm}


\end{document}